\documentclass[12pt,a4paper]{article}

\usepackage{amsmath}
\usepackage{caption}
\usepackage{subcaption}
\usepackage{indent first}
\usepackage{jheppub}
\usepackage{pdfcomment}
\usepackage{tikz}
\usetikzlibrary{shapes,matrix}

\newcommand{\bC}{\mathbb{C}}
\newcommand{\bR}{\mathbb{R}}
\newcommand{\bZ}{\mathbb{Z}}
\def\Nequals#1{$\mathcal{N}{=}#1$}
\def\SU{\mathrm{SU}}

\def\USp{\mathrm{USp}}
\def\U{\mathrm{U}}
\def\SO{\mathrm{SO}}
\def\tr{\mathop{\mathrm{tr}}\nolimits}

\def\sign{\mathop{\mathrm{sign}}\nolimits}
\def\Tr{\mathop{\mathrm{Tr}}\nolimits}

\def\vev#1{\langle#1\rangle}

\def\beq#1\eeq{\begin{align}#1\end{align}}

\usepackage{xcolor}

\begin{document}

\title{Anomaly polynomial of general 6d SCFTs}
\abstract{We describe a method to determine the anomaly polynomials of general 6d \Nequals{(2,0)} and \Nequals{(1,0)} SCFTs, in terms of the anomaly matching on their tensor branches. This method is almost purely field theoretical, and can be applied to all known 6d SCFTs. We demonstrate our method in many concrete examples, including \Nequals{(2,0)} theories of arbitrary type and the theories on M5 branes on ALE singularities, reproducing the $N^3$ behavior. We check the results against the anomaly polynomials computed M-theoretically via the anomaly inflow. 
}
\author[1]{Kantaro Ohmori,}
\author[1]{Hiroyuki Shimizu,}
\author[1,2]{Yuji Tachikawa,}
\author[3]{and Kazuya Yonekura}
\affiliation[1]{Department of Physics, Faculty of Science, \\
 University of Tokyo,  Bunkyo-ku, Tokyo 133-0022, Japan}
\affiliation[2]{Institute for the Physics and Mathematics of the Universe, \\
 University of Tokyo,  Kashiwa, Chiba 277-8583, Japan}
\affiliation[3]{School of Natural Sciences, Institute for Advanced Study,\\
Princeton, NJ 08540, United States of America}
\preprint{IPMU-14-0285, UT-14-37}

\maketitle

\section{Introduction}\label{Introduction}
In the past few years,  6d \Nequals{(2,0)} superconformal theories have been used effectively as a way to organize and understand various features of lower dimensional supersymmetric dynamics. 
We might hope that similar development with 6d \Nequals{(1,0)} theories is not entirely out of reach.
To orient ourselves, we would like to start by understanding better the properties of 6d theories themselves.  

Let us quickly recall known 6d \Nequals{(1,0)} theories in the literature:
The \Nequals{(2,0)} theories, with the ADE classification, were introduced in \cite{Witten:1995zh,Strominger:1995ac}: they are of course \Nequals{(1,0)} theories. 
The E-string theories are obtained by putting M5-branes within the end-of-the-world $E_8$ brane \cite{Ganor:1996mu,Seiberg:1996vs}.
In \cite{Seiberg:1996qx,Bershadsky:1997sb} theories were found that become gauge theories on their tensor branch.
 M5-branes can also be put on the ALE singularity, with or without the end-of-the-world $E_8$ brane.  
Another method is to consider coincident D5-branes in type IIB or type I theory on top of the ALE singularity \cite{Intriligator:1997kq,Blum:1997mm}. 
The theories discussed so far can be uniformly analyzed  in terms of F-theory \cite{Aspinwall:1997ye}; the brane construction with D6 branes and NS5 branes can also be used \cite{Brunner:1997gf,Hanany:1997gh}.
F-theory gives a uniform perspective to discuss these theories: the classification was started in \cite{Heckman:2013pva} and the details are being worked out, e.g.~\cite{DelZotto:2014hpa,Heckman:2014qba}.\footnote{There are also approaches to study  6d \Nequals{(1,0)} superconformal theories using Lagrangian descriptions, see e.g.~\cite{Ivanov:2005qf,Ivanov:2005kz,Smilga:2006ax} and \cite{Samtleben:2011fj,Samtleben:2012fb}.}

One feature of these 6d superconformal theories is that they have the tensor branch, i.e.~the moduli space of vacua parameterized by the scalars in the tensor multiplets. On the tensor branch, the infrared theory is simpler and described by a system of almost free tensor multiplets, gauge fields and matter contents that can either be free hypermultiplets or other superconformal field theories.  
The scalars in the tensor multiplets often control the coupling constant of the non-Abelian gauge multiplets. 
The objective of this paper is to show how this feature can be used to determine the anomaly polynomial of the original ultraviolet theory, providing us at least one additional physical observable for each 6d superconformal theory. 

The essential idea is that going to the tensor branch does not break any symmetry other than the conformal symmetry. Therefore, the whole anomaly of the ultraviolet theory can be found on the tensor branch by the anomaly matching. The anomaly there has two sources: the one-loop anomaly and the Green-Schwarz contribution.\footnote{This essential idea, of the anomaly matching on the tensor branch, was independently found earlier by Ken Intriligator, and it appeared on the arXiv as \cite{Intriligator:2014eaa}.}  The one-loop anomaly follows from the standard formulas, and therefore all we need to do is to determine the Green-Schwarz contribution, which can be found in either of the two methods: \begin{enumerate}
\item If there is no gauge group whose coupling is controlled by the tensor multiplet scalar, we compactify the system on $S^1$, determine the Chern-Simons term in 5d, which can be lifted back to 6d.
\item If there is a gauge group whose coupling is controlled by  the tensor multiplet scalar, the requirement of the cancellation of the gauge anomaly uniquely fixes the Green-Schwarz term. 
\end{enumerate}
These methods allow us, in particular, to derive the characteristic $N^3$ behavior of the number of the degrees of freedom on 6d superconformal theories in an almost purely field theoretical manner. 
We think it best to demonstrate our methods using a few concrete examples here.

The R-symmetry of 6d \Nequals{(1,0)} theories is $\SU(2)_R$, and in the Introduction, 
we are going  to determine the $c_2(R)^2$ term in the anomaly polynomial of a few typical \Nequals{(1,0)} theories, 
where $c_2(R)=\Tr F_R^2/4$
is the second Chern class of the background $\SU(2)_R$-symmetry bundle.
(Throughout the paper, a factor of $1/2\pi $ is included in field strengths $F$, and $\Tr$ denotes the trace in the adjoint representation divided by the dual Coxeter number. Therefore, integral of $\frac{1}{4} \Tr F^2$ gives the instanton number.)
During the Introduction, we only include terms involving $c_2(R)$ and $\Tr F^2$ for the gauge fields in the anomaly polynomial 
to keep the expressions briefer.

\subsection{Without 6d gauge group}
Let us consider the anomaly of the  low-energy theory on $Q$ M5-branes, which is the 6d \Nequals{(2,0)} theories of type $A_{Q-1}$. 
There is a long history behind  the computation of the anomaly polynomials of these theories, using M-theoretic techniques.  
For a single M5-brane, it was first discussed in \cite{Duff:1995wd}. The anomaly inflow analysis for a single M5-brane was done in \cite{Witten:1996hc}, where a subtlety concerning the normal bundle anomaly was found. This subtlety was successfully resolved in \cite{Freed:1998tg}, which led to the determination of the anomaly for general number $Q$ of the M5-branes in \cite{Harvey:1998bx}.

Our trick is to go to its \Nequals{(1,0)} tensor branch.
On generic points on the tensor branch, we just have $Q$ \Nequals{(1,0)} tensor multiplets and $Q$ hypermultiplets, whose contribution to the $c_2(R)^2$ term in the anomaly polynomial is  just \begin{equation}
I^\text{one-loop}=\frac{Q}{24}c_2(R)^2.\label{intro:oneloop:m5}
\end{equation}
Going to the \Nequals{(1,0)} branch, however, does not break the $\SU(2)_R$  symmetry. Therefore, we should be able to see the full $\SU(2)_R$ anomaly of the interacting theory  on the tensor branch: it should have additional contribution from the Green-Schwarz term. Namely, if the $i$-th tensor field has the modification of the Bianchi identity as \begin{equation}
dH_i=I_i = \rho_i c_2(R) 
\end{equation} the Green-Schwarz contribution to the anomaly is \cite{Sadov:1996zm,Kumar:2010ru,Monnier:2013kna} \begin{equation}
I^\text{GS}=\frac12 \delta^{ij} I_i I_j = \frac12 |\rho|^2 c_2(R)^2. \label{intro:gs:m5}
\end{equation} We just need  to  determine $\rho_i$.

To do this, we perform a dimensional reduction on $S^1$.  We have maximally supersymmetric $\U(Q)$ theory on 5d. The \Nequals{(1,0)} tensor branch in 6d corresponds to giving the vev to only one direction of the scalars out of $\bR^5$, breaking $\U(Q)$ down to $\U(1)^Q$ gauge group. Let us say that the vev is \begin{equation}
\phi_1 < \phi_2 < \cdots < \phi_Q.
\end{equation}  We have corresponding $\U(1)$ gauge fields $A_{i=1,\ldots,Q}$. 
For each pair $(i,j)$ with $i\neq j$, we have massive vector multiplets with real mass $\phi_i-\phi_j$,
with charge $+1$ for $A_i$ and charge $-1$ for $A_j$. 
In five dimensions, integrating  out fermions in these massive multiplets generates Chern-Simons interactions. 
We are interested in $A_i$-$F_R$-$F_R$ Chern-Simons terms, where $F_R$ is the background gauge field for the $\SU(2)_R$ symmetry.  A multiplet with $\U(1)$ charge $q$ and real mass $m$ gives a contribution proportional to $q \sign m$. 
Under the $i$-th $\U(1)$ field $F_i$, the pairs $(i,j)$ all have charge $+1$, but those with $j>i$ have positive real mass, whereas those with $j<i$ have negative real mass. Therefore, we see
\begin{equation}
d \star F_i \propto [(Q-i) - (i-1)] \Tr F_R^2.
\end{equation} Lifting it and carefully fixing the coefficients, which we will do in Sec.~\ref{sec:without}, we find \begin{equation}
dH_i = \frac12(Q+1-2i) c_2(R).
\end{equation}  This determines the Green-Schwartz contribution $I^\text{GS}$ of \eqref{intro:gs:m5}, and we get \begin{equation}
I^\text{tot} = I^\text{one-loop} + I^\text{GS} = \frac{Q^3}{24} c_2(R)^2.
\end{equation}

This correctly reproduces the $Q^3$ behavior of the anomaly polynomial including the coefficients.
In addition, this procedure applies equally well to 6d \Nequals{(2,0)} theories of any type $G=A_n$, $D_n$ and $E_n$. The general formula was conjectured in \cite{Intriligator:2000eq}, and the anomaly polynomial of \Nequals{(2,0)} theory of $D_n$ was obtained by inflow analysis in \cite{Yi:2001bz}.\footnote{In \cite{Maxfield:2012aw}, compactification on $T^2$ and comparison with \Nequals{4} super Yang-Mills in 4d were used to deduce the $N^3$ behavior of the anomaly of 6d \Nequals{(2,0)} theory of type $A$ and $D$ in the large $N$ limit.}
There has been, however, no derivation for the theory of type $E$.
Our method gives the first derivation of the anomaly polynomial of the 6d theory of type $E$. 

We will present the details of the procedure described in this subsection in Sec.~\ref{sec:without}.
We will treat the \Nequals{(2,0)} theories and E-string theories there.

\subsection{With 6d gauge group}
\subsubsection{Rank-1 examples}
Let us next consider the class of 6d \Nequals{(1,0)} theories whose tensor branch is  one dimensional, such that on its generic point we just have pure gauge theory with gauge group $G=\SU(3)$, $\SO(8)$,  $F_4$, $E_{6,7,8}$.   These theories were first introduced in \cite{Seiberg:1996qx,Bershadsky:1997sb}.

The anomaly polynomial of the gauge multiplet is \begin{equation}
I^\text{vec}= -\frac1{24}( \frac34 w_G (\Tr F^2)^2 + 6h^\vee_G \Tr F^2 c_2(R)+ d_G c_2(R)^2).\label{intro:gauge}
\end{equation} where $3w_G/4$ is the coefficient converting $\tr_{\rm adj} F^4$ to $(\Tr F^2)^2$,  $h^\vee_G$ and $d_G$ are the dual Coxeter number and the dimension of $G$, respectively. 
These and other data and also our convention are collected in Appendix~\ref{sec:tables}.
  For simplicity we only showed the terms that only involve the gauge fields and the R-symmetry.
The one-loop anomaly on the tensor branch is then \begin{equation}
I^\text{one-loop}=I^\text{vec}+I^\text{tensor} 
\end{equation} where $I^\text{tensor}=\frac1{24}c_2(R)^2$ is the contribution from the tensor multiplet. 
The self-dual tensor field can have a deformation of the Bianchi identity $dH=I$
where $I$ is a linear combination of $\Tr F^2$ and $c_2(R)$. 
Depending on the normalization of $H$, it contributes to the anomaly by $I^{GS}=a  I^2$ where $a$ is a positive number.  
To cancel the pure and mixed gauge anomalies in \eqref{intro:gauge}, the essentially unique choice is to take \begin{equation}
	I^\text{GS}=\frac{w_G}{2}(\frac14\Tr F^2+\frac{h^\vee_G}{w_G} c_2(R))^2. \label{eq:rank1}
\end{equation}  We conclude that the total anomaly polynomial is \begin{equation}
I^\text{tot}=I^\text{one-loop}+I^\text{GS}=\left(\frac{(h^\vee_G)^2}{2w_G}-\frac{d_G-1}{24}\right) c_2(R)^2.
\end{equation}
Note that this is the anomaly polynomial of the ultraviolet fixed point. This is explicit and concrete, but not very illuminating. Let us move on to another class of examples. 

\subsubsection{$Q$ M5-branes on $\bC^2/\bZ_k$}

Consider $Q$ M5-branes on the singularity $\bC^2/\bZ_k$, without the center of mass mode.  
The tensor branch is $(Q-1)$-dimensional, and on its generic point, the theory is a linear quiver theory $[\SU(k)_0]\times \SU(k)_1\times \cdots \times \SU(k)_{Q-1} \times [\SU(k)_Q]$ with $(Q-1)$ gauge factors $\SU(k)_{1,\ldots,Q-1}$ and flavor symmetry $\SU(k)_0\times \SU(k)_Q$, with bifundamental hypermultiplets under $\SU(k)_i\times \SU(k)_{i+1}$. These theories were first considered in \cite{Intriligator:1997kq} and studied using various stringy constructions in \cite{Blum:1997mm,Brunner:1997gf,Hanany:1997gh}.

 Let us determine the anomaly polynomial of this strongly-coupled theory. 
The anomaly polynomial of the gauge multiplet for $\SU(k)_i$ is \begin{equation}
I^\text{vec}_i= -\frac1{24}(2k \tr_{\rm fund} F_i^4 + (3/2) (\Tr F_i^2)^2 + 6k\Tr F_i^2 c_2(R)+(k^2-1)c_2(R)^2).
\end{equation} 
Similarly, the anomaly of the bifundamental charged under $\SU(k)_i\times \SU(k)_{i+1}$ is \begin{equation}
I^\text{bif}_{i,i+1}=\frac1{24}(k \tr_{\rm fund}  F_i^4+k \tr_{\rm fund}  F_{i+1}^4 + \frac32 \Tr F_i^2 \Tr F_{i+1}^2 )
\end{equation}   and that of one tensor multiplet is \begin{equation}
I^\text{tensor}=\frac1{24}{c_2(R)^2}.
\end{equation} The contribution so far sums up to \begin{equation}
I^\text{one-loop}= -\frac1{32} \eta^{ij} \Tr F_i^2 \Tr F_j^2 -\frac k4 \Tr F_i^2 \rho^i c_2(R) -\frac{1}{24}(Q-1)(k^2-2) c_2(R)^2 \label{oneloop}
\end{equation} where $\eta^{ij}$ for $i,j=1,\ldots,Q-1$ is given by the Cartan matrix of $A_{Q-1}$ that is  \begin{equation}
\eta^{ij}=\begin{pmatrix}
2 &- 1 & \\
-1 & 2 & -1 \\ 
& \ddots & \ddots & \ddots \\
&&-1&2&-1\\
&&&-1 & 2
\end{pmatrix},
\end{equation} the vector $\rho$ is \begin{equation}
\rho^i=(1,1,\ldots,1)
\end{equation} and for simplicity we set the flavor background to be zero: $F_0=F_Q=0$.

This gauge theory is consistent only because there are $(Q-1)$ self-dual tensor fields whose Green-Schwarz interaction cancels the purely and mixed gauge anomalies. 
In general, the Green-Schwarz contribution from the self-dual tensor fields to the anomaly polynomial is \begin{equation}
I^\text{GS}=\frac12\Omega^{ij} I_i I_j
\end{equation}  where $\Omega^{ij}$ is a positive-definite matrix and $I_i$ is the modification to the Bianchi identity for the $i$-th self-dual field \begin{equation}
dH_i = I_i.
\end{equation}
Here, the essentially unique choice to cancel the gauge anomaly in \eqref{oneloop} is to take \begin{equation}
\Omega^{ij}=\eta^{ij},\qquad I_i = \frac14\Tr F_i^2 + k (\eta^{-1})_{ij} \rho^j c_2(R) .
\end{equation} Then we conclude \begin{align}
I^\text{tot}=I^\text{one-loop}+I^\text{GS}&=
(\frac{k^2}2 \rho^i(\eta^{-1})_{ij} \rho^j -\frac1{24}(Q-1)(k^2-2))   c_2(R)^2 \nonumber\\
&= \frac1{24}( (Q^3-Q)k^2 - (Q-1) (k^2-2) )c_2(R)^2
\end{align}
where we used $\rho^i(\eta^{-1})_{ij} \rho^j =(Q^3-Q)/12$.

We already see that  this purely field theoretical method already gives the leading cubic behavior $Q^3$.
The coefficient is exactly what is expected from AdS$_7$/CFT$_6$; the whole structure can also be obtained from the anomaly inflow in M-theory, see Appendix~\ref{sec:inflow}.

\subsection{Organization of the paper}
The rest of the paper is organized as follows. 
In Sec.~\ref{sec:without}, we give the details of the determination of the Green-Schwarz term when the tensor multiplet scalar does not control the coupling of any gauge field. 
It happens that in this particular case, we have a good control over the field theoretical behavior of its $S^1$ compactification to 5d, that allows us to determine the 5d Chern-Simons terms appearing from integrating out massive fermions. 
Concretely, we treat 6d \Nequals{(2,0)} theories of arbitrary type $G=A_n$, $D_n$ and $E_n$, and also the E-string theory of arbitrary rank.  Our results reproduce the known anomaly polynomials computed using the anomaly inflow in M-theory. Note that this is the first time where the anomaly of \Nequals{(2,0)} theory of type $E$ was successfully computed. 

In Sec.~\ref{sec:with}, we describe the methods to find the Green-Schwarz term when the scalar in the tensor multiplet determines the coupling of a gauge field. Here, the varieties of 6d \Nequals{(1,0)} theories we can treat is vast. We will treat M5-branes on $\bC^2/\Gamma$ for arbitrary $\Gamma$ as the main examples.
Most of the results we obtain in this section are new. 
At various steps, one needs to use the anomaly polynomials of the E-string theories as inputs.
We end the section by re-phrasing our results in terms of the F-theory geometry, used in the classification in \cite{Heckman:2013pva}.

We conclude by a discussion of future directions in Sec.~\ref{sec:conclusions}.

We have two appendices: in Appendix.~\ref{sec:tables}, we gather various standard formulas, such as the anomaly polynomials of various \Nequals{(1,0)} multiplets and the group theoretical constants. 
In Appendix.~\ref{sec:inflow}, we generalize the anomaly inflow analysis of \cite{Freed:1998tg, Harvey:1998bx} to determine the anomaly polynomials of M5-branes on $\bC^2/\Gamma$. This gives an independent confirmation of our methods in Sec.~\ref{sec:with}. 

\section{Tensor branches without gauge multiplets}\label{sec:without}
In this section we determine the anomaly polynomials of \Nequals{(2,0)} theories of arbitrary type
and of  E-string theories of arbitrary rank, by going to their tensor branches. 
If the reader accepts the anomaly polynomials of these theories as known from the M-theoretic anomaly inflow, the content of this section is not necessary and the reader can directly go to the next section. 

\subsection{Generalities of self-dual tensor fields in 6d}\label{sec:generality}
We start by recalling various properties of self-dual tensor fields in 6d. 
Let us first introduce the charge pairing in 6d. Before that, it is useful to recall the situation in 4d.
We normalize the 4d Dirac-Zwanziger pairing of particles with dyonic charges $q=(e,m)$ and $q'=(e',m')$ to be given by \begin{equation}
\vev{q,q'}_{4d}=em'-e'm \in \bZ
\end{equation} so that $\vev{q,q'}\hbar/2$ gives the angular momentum carried by the electromagnetic fields. This pairing is anti-symmetric. 

In 6d with $n$ self-dual tensor fields, there are self-dual strings with charges taking values in an $n$ dimensional lattice $\Lambda$.  The pairing is symmetric: for $q,q'\in \Lambda$, $\vev{q,q'}_{6d}=\vev{q',q}_{6d}$, and we normalize $\vev{q,q'}_{6d}$ using the compactification on $T^2$. 
Namely,  the self-dual string of charge $q$ wound on the cycle $mA+nB$ of $T^2$ can be said to have 4d charge $q(mA+nB)$, and we require \begin{equation}
\vev{qA,q'B}_{4d}=\vev{q,q'}_{6d}\vev{A,B}_{T^2}
\end{equation} where $\vev{A,B}_{T^2}$ is the intersection number of $A$ and $B$ on $T^2$.

Let us explicitly introduce $q=(q_i)_{i=1,\ldots,n} \in \Lambda$, and express the pairing using a symmetric matrix $\Omega^{ij}$ as \begin{equation}
\vev{q,q'}_{6d}=\Omega^{ij} q_i q'_j.\label{6dpairing}
\end{equation} Accordingly, introduce the self-dual three-form field strengths $H_i$ normalized such that  \begin{equation}
dH_i = q_i \prod_{a=2,3,4,5}\delta(x_a)dx_a
\end{equation} when a self-dual string of charge $q$ exists at $x_{a=2,3,4,5}=0$. 

At this point, suppose we have a modification of the Bianchi identity \begin{equation}
dH_i = I_i\label{bianchimod}
\end{equation} where $I_i$ is a four-form constructed out of the metric and the gauge fields, which can either be dynamical or non-dynamical.  The Green-Schwarz contribution to the anomaly is \cite{Sadov:1996zm,Kumar:2010ru,Monnier:2013kna} \begin{equation}
I^\text{GS}=\frac12 \Omega^{ij} I_i I_j.
\end{equation}

\subsection{6d Green-Schwarz and 5d Chern-Simons}
Next, we recall the relation of the 6d Green-Schwarz terms and the 5d Chern-Simons terms,
and also how Chern-Simons terms are induced in 5d.
The $S^1$ reduction of $n$ self-dual fields $H_i$ gives rise to $n$ Abelian gauge fields $A_i$. 
The field strengths are related as $F_{\mu\nu}=2 \pi R \cdot H_{\mu\nu 5 } $, where $R$ is the radius of $S^1$ and $``5"$ is the direction of $S^1$.
The 5d kinetic term is $\frac{1}{2R} \Omega^{ij} F_i\wedge \star F_j$, 
and the reduction of \eqref{bianchimod} is
 \begin{equation}
d( \frac{1}{2\pi R} \star F_i) = I_i,
\end{equation} meaning that there is a 5d Chern-Simons term\footnote{Our normalization of $p$-form fields (including gauge fields) are such that 
their field strengths take values in integer cohomology $H^{p+1}(M, \bZ)$ for a smooth manifold $M$. This makes the normalization of gauge fields 
to be different by a factor of $2\pi$ from the usual one.}
\begin{equation}
\frac{1}{2\pi }S^\text{CS}= \Omega^{ij} A_i I_j=A_i I^i,
\end{equation}
where indices are raised and lowered by $\Omega^{ij}$ and $(\Omega^{-1})_{ij}$, e.g. $I^i = \Omega^{ij} I_j$.

Now, let us consider a 5d fermion $\psi$ with mass term $m\psi\psi$; note that the sign of $m$ is meaningful.  Suppose it has charge $q$ under a $\U(1)$ field, and furthermore, it couples to an additional non-Abelian background gauge field $F_G$ in a representation $\rho$ and of course the metric. By a careful computation of the triangle diagrams \cite{Bonetti:2013ela}, the induced Chern-Simons term from integrating out $\psi$ is  
\begin{equation}
\frac12 (\sign m) q A (\frac12\tr_\rho F_G^2 + \frac1{24} d_\rho p_1(T)). \label{inducedCS}
\end{equation} 

This result can also be obtained as follows. Consider we have a time-translation invariant situation where we have nontrivial $F_G$ and/or nontrivial metric on the spatial slice. In this background, the fermion $\psi$ has $\nu = -\tr_\rho F^2/2 -  d_{\rho} p_1(T)/24$ zero modes. By quantizing the fermionic zero modes in the instanton background, we can see that each zero mode shifts the electric charge by $\pm q/2$, depending on the sign of $m$. Then the worldline Lagrangian for $\psi$ has an additional coupling $\pm (q\nu/2) A$, which indicates the 5d Chern-Simons term \eqref{inducedCS}.

\subsection{\Nequals{(2,0)} theories}
Now we have all the tools to compute the anomaly polynomial of 6d \Nequals{(2,0)} theory of arbitrary type $G=A_r$, $D_r$ and $E_r$. Its $T^2$ reduction is 4d \Nequals4 theory with gauge group $G$.  Therefore, the 6d charge lattice of the self dual strings is the root lattice of $G$, and the matrix $\Omega^{ij}$ in \eqref{6dpairing} is the Cartan matrix $\eta^{ij}$ of type $G$.

The R-symmetry of the \Nequals{(2,0)} theory is $\SO(5)_R$. As an \Nequals{(1,0)} theory, we can see the symmetry $\SU(2)_R \times \SU(2)_L\simeq \SO(4)_R \subset \SO(5)_R$. $\SU(2)_L$ is a flavor symmetry from the \Nequals{(1,0)} viewpoint. Going to the \Nequals{(1,0)} tensor branch does not break $\SO(4)_R$. 

Now, reduce the system on $S^1$.  The \Nequals{(1,0)} tensor branch corresponds to giving vevs to only one direction out of $\bR^5$ worth of scalars.  We consider a generic vev  $v\in \mathfrak{h}$, where $\mathfrak{h}$ is the Cartan subalgebra of $\mathfrak{g}$, such that the low energy system is just $\U(1)^r$.  For each root $\alpha\in\mathfrak{h}$, we have a massive charged \Nequals2 vector multiplet of mass $|v\cdot \alpha|$, i.e. a pair of a massive \Nequals1 vector multiplet and a massive \Nequals1 hypermultiplet. Note that the fermion mass term is of the form \begin{equation}
\psi \Gamma^I \phi^I \psi
\end{equation} where $\psi$ is in the spinor representation of $\SO(5)_R$ and $I=1,\ldots,5$ is the index for the vector representation of $\SO(5)_R$.  We are giving the vev to only $\phi^{I=5}$. 
Therefore, the \Nequals1 vector multiplet is charged only under $\SU(2)_R$ and has the real mass $-v\cdot \alpha$, whereas the \Nequals1 hypermultiplet is charged only under $\SU(2)_L$ and has the real mass $+v\cdot \alpha$. 

We can use $v$ to determine the positive side and the negative side of the Cartan subalgebra; accordingly, the roots $\alpha$ can be separated into the positive roots and the negative roots.  The induced Chern-Simons terms for the $\U(1)^r$ field $A$ valued in $\mathfrak{h}$ is then \begin{equation}
\frac12 \sum_{\alpha>0} (\alpha \cdot A) \left[(c_2(L)+\frac2{24}p_1(T))-(c_2(R)+\frac2{24}p_1(T))\right]
=\rho\cdot A (c_2(L)-c_2(R))
\end{equation} where $\rho$ is the Weyl vector.   Lifting it back to 6d, the Bianchi identity of the self-dual tensor fields is given by \begin{equation}
dH_i = \rho_i (c_2(L)-c_2(R)),
\end{equation}
and therefore, the Green-Schwarz contribution to the anomaly of the 6d theory is \begin{equation}
\frac12\vev{\rho,\rho} (c_2(L)-c_2(R))^2 = \frac{h^\vee_G d_G}{24} (c_2(L)-c_2(R))^2
\end{equation} where we used the strange formula of Freudenthal and de Vries.

We conclude that the anomaly polynomial of the 6d \Nequals{(2,0)} theory of type $G$ is given by \begin{equation}
I^\text{\Nequals{(2,0)}}_G=  \frac{h^\vee_G d_G}{24} p_2(N) +  r_G I^\text{\Nequals{(2,0)} tensor},
\end{equation} where we added the contribution from $r_G$ free \Nequals{(2,0)} tensor multiplets on the tensor branch, and used the fact that $\chi_4(N)=c_2(L)-c_2(R)$, $p_2(N)=\chi_4(N)^2$ when the $\SO(5)_R$ bundle is in fact an $\SO(4)_R\simeq \SU(2)_L\times \SU(2)_R$ bundle. 

The anomaly of $Q$ M5-branes is obtained by adding one additional free \Nequals{(2,0)} tensor to the 6d \Nequals{(2,0)} theory of type $A_{Q-1}$, and has the form \begin{equation}
I^\text{$Q$ M5s}=\frac{Q^3}{24}p_2(N) - Q I_8
\end{equation} where 
\begin{equation}
I_8=\frac1{48}(p_2(N)+p_2(T)-\frac14(p_1(N)-p_1(T))^2).
\end{equation}

\subsection{E-string theory of arbitrary rank}\label{sec:Est}
Next, let us consider the E-string theory of rank $Q$. This is the low-energy theory on $Q$ M5-branes on top of the end-of-the-world $E_8$ brane. For now, let us include the free hypermultiplet corresponding to the motion parallel to the $E_8$ brane. 

We use the fact that when it is put on $S^1$ with a holonomy of $E_8$ chosen so that it breaks $E_8$ to $\SO(16)$, the 5d theory is given by \Nequals1 $\USp(2Q)$ theory with an antisymmetric hypermultiplet and 8 hypermultiplets in the fundamental representation.   This allows us to reconstruct the full anomaly polynomial, since $\SO(16)$ is a maximal rank subgroup of $E_8$.

We can go to the generic point on the tensor branch and repeat the analysis as in the case of \Nequals{(2,0)} theories. 
Instead, let us consider a rather special point on the  tensor branch such that $Q$ M5-branes are still coincident but is separated from the end-of-the-world brane.  
There is only one tensor multiplet scalar activated, which is a diagonal sum of $Q$ free tensor multiplets on the generic points. 
Therefore, the matrix $\Omega^{ij}$ in \eqref{6dpairing} is $Q$. 

In 5d, this point on the Coulomb branch corresponds to giving a vev to the adjoint scalar of the vector multiplet which 
breaks $\USp(2Q)$ to $\U(Q)$; we would like to determine the Chern-Simons term involving the $\U(1)$ part.  We have $Q$ massive hypermultiplets in the vector representation of $\SO(16)$ with $\U(1)$ charge 1, and  $Q^2+Q$ massive vector multiplets and $Q^2-Q$ hypermultiplets, both with $\U(1)$ charge 2.  Recalling that the vector is charged under the $\SU(2)_R$ and the hypers in the anti-symmetric representation
under the $\SU(2)_L$, we find the induced Chern-Simons term to be \begin{multline}
\frac12 A \wedge \Bigl[
Q(\frac{\Tr F^2}2+\frac{16p_1(T)}{24})+
2(Q^2-Q) \frac12(c_2(L)+\frac{2p_1(T)}{24}) \\
-2(Q^2+Q) \frac12(c_2(R)+\frac{2p_1(T)}{24})
\Bigr] =\Omega A\wedge (\frac{Q}2\chi_4(N) +  I_4)
\end{multline} where $\Omega=Q$ and \begin{equation}
I_4= \frac14( \Tr F^2+ p_1(T) + p_1(N))
\end{equation}  and we again used \begin{equation}
\chi_4(N)=c_2(L)-c_2(R), \qquad 
p_1(N)=-2(c_2(L)+c_2(R)).
\end{equation} From this, we find that \begin{align}
I^\text{E-string, rank $Q$+free hyper} 
&= I^\text{$Q$ M5s} + \frac{Q}{2}  (\frac{Q}2\chi_4(N) +  I_4)^2 \\
&=\frac{Q^3}{6} \chi_4(N)^2 + \frac{Q^2}2 \chi_4(N) I_4 
+ Q(\frac12 I_4^2 - I_8).\label{E-string-anomaly}
\end{align} where $I_8$ was given above. This reproduces the result in \cite{Ohmori:2014pca} obtained via the anomaly inflow. Note that this contains the contribution of a free hypermultiplet with the anomaly \begin{equation}
I_\text{free}=\frac{7p_1(T)^2-4p_2(T)}{5760} +\frac{c_2(L) p_1(T)}{48} + \frac{c_2(L)^2}{24}.\label{freehyper}
\end{equation} When the E string theory is used as a matter content, we always need to subtract this contribution \eqref{freehyper} from \eqref{E-string-anomaly}.

\section{Tensor branches with gauge multiplets}\label{sec:with}
In this section we develop the method to determine the anomaly polynomials of 6d \Nequals{(1,0)} theories when we have non-Abelian gauge fields on the tensor branch.
As briefly explained in Introduction, we can uniquely determine the Green-Schwarz terms by requiring the gauge anomalies vanish.
After explaining the basic ideas, we focus on the case of the worldvolume theories on multiple coincident  M5 branes on ALE singularities of arbitrary type.
We end the section by  explaining the relation between the Green-Schwarz terms and the F-theory construction of arbitrary 6d \Nequals{(1,0)} theories.

\subsection{Basic ideas}\label{sec:ideas}
A 6d \Nequals{(1,0)} superconformal theory can have the tensor branch,
such that 
 the infrared theory at a point on the tensor branch consists of $t$ vector multiplets in gauge group $G_A$, $A=1, \cdots, t$, and $t$ free tensor multiplets whose scalars give the coupling constants of $G_A$,
 together with a number of charged ``bifundamental matter contents''. 
These ``bifundamental matter'' can either be Lagrangian hypermultiplets or another 6d SCFT whose flavor symmetries are gauged by $G_A$. 
We assume that anomalies of the ``bifundamental matter'' are already known.
This is indeed the case for all the theories discussed in \cite{Heckman:2013pva}, where we can have E-string theories of 
 rank one or two as the ``matter contents''.
 
 Note that although the full tensor branch of the theory may have a dimension larger than the number of the gauge groups $t$,
 we always stay on the subspace of the full tensor branch where the number of almost free tensor multiplets is the same as that of the gauge groups.
 In particular, this means that we do not give vevs to the tensor modes of E-string theories.

Now, the ``one-loop" anomaly (i.e. the anomaly without Green-Schwarz contribution) on the tensor branch is given by
\begin{equation}
I^{\rm one-loop}=\sum_{A} I^{\text{vec}}_{F_A} + \sum_{A,B}I^{\text{matter}}_{F_A, F_B} + tI^{\text{tensor}}.
\end{equation}
It contains pure gauge and mixed gauge-background terms,
\begin{equation}
I^{\rm one-loop}\supset -\frac{1}{32}c^{AB} \Tr F^2_A \Tr F^2_B -\frac{1}{4}X^A \Tr F^2_A,
\end{equation}
where $X^A$ consists of only background flavor and gravity fields.
One needs to cancel these gauge anomalies by the Green-Schwarz contribution,
\beq
\frac{1}{2}\Omega^{ij} I_i I_j.
\eeq
Here $\Omega^{ij}$ is the symmetric matrix introduced in \eqref{6dpairing} which, roughly speaking, is the matrix appearing in the kinetic term
of the tensor multiplets. 
The anomaly cancellation requires,
\begin{align}	
	I_i &= \frac{1}{4}d_{i}^A\Tr F^2_A +(\Omega^{-1})_{ij} (d^{-1})_A^j  X^A ,~~~~~ d_i^A d_j^B \Omega^{ij}=c^{AB}.
\end{align}
where we have assumed that the matrix $c^{AB}$ has the maximal rank $t$ which implies that the matrix $d_i^A$ is invertible.
This is the point we need the number of free tensor multiplets and the gauge groups $G_A$ to be the same.

Although the matrix $d_i^A$ is not completely determined, the Green-Schwarz contribution is uniquely determined 
in terms of $c^{AB}$ and $X^A$ as
\beq
\frac{1}{2}\Omega^{ij} I_i I_j=\frac{1}{32}c^{AB} \Tr F^2_A \Tr F^2_B +\frac{1}{4}X^A \Tr F^2_A+\frac{1}{2}(c^{-1})_{AB}X^A X^B.
\eeq
The first two terms cancel the gauge anomalies, and the third term gives the Green-Schwarz contribution to the anomaly of background fields.

\subsubsection{A consistency condition on the theory}
Before going to the applications of the above method, we would like to make an interesting digression here. 
The self-dual three-form field strengths $H_i$ in the tensor multiplets satisfy $dH_i=I_i$ as discussed in Sec.~\ref{sec:generality}.
From the above expression for $I_i$, a point-like instanton of the gauge field $\frac{1}{4}\Tr F^2_A$ gives a string in 6d with charge $q_i =d^A_i$.
Then, each element of $c^{AB}=\Omega^{ij} d_i^A d_j^B=\vev{d^A,d^B}_{6d}$ must be an integer precisely by the 6d charge quantization condition 
discussed around \eqref{6dpairing}. This imposes a strong constraint on the theory.
For example, the theories which are excluded based on global anomalies in \cite{Bershadsky:1997sb}
can already be excluded by this constraint alone, because the one-loop anomaly polynomial yields non-integer $c^{AB}$ in those theories.

\subsection{M5 branes on ALE singularities}
As an example of the method outlined in Sec.~\ref{sec:ideas}, we determine the anomalies of $Q$ M5 branes on an ALE singularity $\bC^2/\Gamma$.
When $\Gamma$ is of type $A_k$, there is a $\U(1)$ symmetry acting on $\mathbb{C}^2/\Gamma$, but we ignore this symmetry for simplicity.

In M-theory, the singular locus extends along seven dimensions, on which lives a 7d dynamical gauge multiplet in the gauge group $G$ determined by $\Gamma$. 
M5-branes are six dimensional, and therefore we consider the singular locus to form a line of singularities transverse to the worldvolume. 

We first separate $Q$ M5 branes along this line of singularities. 
The system can be described as a linear quiver theory \begin{equation}
	[G_0]\times G_1\times \cdots \times G_{Q-1} \times [G_Q]\label{Gquiver}
\end{equation}
 with $(Q-1)$ gauge factors $G_{1,\ldots,Q-1}$ and flavor symmetry $G_0\times G_Q$, and ``conformal matters'' charged under $G_i \times G_{i+1}$. The ``conformal matter'' is a theory which is realized on a single M5 brane on the singularity. 
So,  the computation of the anomaly of $Q$ M5-branes involves two steps. The first  is to compute the anomalies of each ``conformal matters''.
 The second is to compute the anomalies of the linear quiver theory.

 \subsubsection{Anomalies of ``conformal matters''}
The ``conformal matter'' is a Lagrangian hypermultiplet when $G$ is of type $A$, and another nontrivial 6d SCFT when $G$ is of type $D$ or $E$. Since we already know anomalies of Lagrangian hypermultiplets, we concentrate on the latter.  

The tensor branch of these SCFTs is investigated in \cite{DelZotto:2014hpa}.
What happens is that a single M5 brane can split to several fractional M5 branes along the line of singularities. On a generic point on the tensor branch, the low energy theory consists only of tensor multiplets, hypermultiplets and vector multiplets. 
The tensor multiplet scalars are the distances between two adjacent fractional M5-branes.
However, it is not always that there is a nontrivial gauge group on a segment between two fractional M5-branes.  If this happens, we make these fractional M5-branes coalesce. 
Then we have the situation where there are an equal number of tensor and vector multiplets, coupled to Lagrangian matter fields and/or E-string theories of rank 1 and 2.  
Then, we can just apply the method in Sec.~\ref{sec:ideas}.

Let us list the structure of the ``conformal matters'' at the point on the tensor branch we use to identify the anomaly polynomial, following \cite{DelZotto:2014hpa}. We will use the $E_6$ case to illustrate detailed steps of the computation.

\paragraph{($A_{k-1}$, $A_{k-1}$) conformal matter.}
This is just a hypermultiplet in the bifundamental of $\SU(k)\times \SU(k)$.

\paragraph{($D_{k}$, $D_{k}$) conformal matter.}
The tensor branch is one dimensional, or equivalently, the M5-brane can fractionate into two.
The first fractional M5-brane changes the gauge group from $\SO(2k)$ to $\USp(2k-8)$.
The second fractional M5-brane changes it back to $\SO(2k)$.  We can depict the setup 
\begin{equation}
\SO(2k) | \USp(2k-8)  | \SO(2k)
\end{equation} where $|$ stands for a fractional M5-brane, and the groups displayed are the gauge groups on the particular half-line or segment of the $D_k$ singularities.  One can also regard it as describing the linear quiver gauge theory, where the two $\SO(2k)$ at the ends are flavor symmetries, and $\USp(2k-8)$ is a gauge symmetry.  The fractional M5-brane between $\SO(2k)$ and $\USp(2k-8)$ provides a half-hypermultiplet in the bifundamental. 

In the case $k=4$, there is no $\USp$ gauge group between two fractional M5 branes, so our method cannot be applied.
The conformal matter realized on one full M5 brane on $D_4$ is actually the rank-1 E-string theory.
However, the anomaly polynomial of this theory is also given by putting $k=4$ in the general formula we will present later.
 
\paragraph{($E_6$, $E_6$) conformal matter.}
The tensor branch is three dimensional, and the M5-brane can fractionate into four.  The gauge groups that occur between the fractional M5-branes are \begin{equation}
E_6 |  \text{empty} | \SU(3) | \text{empty} | E_6.
\end{equation} To compute the anomaly, we make pairs of fractional M5-branes to coalesce: \begin{equation}
E_6 || \SU(3) || E_6.
\end{equation} Now we have a SU(3) vector multiplet plus one $(1,0)$ tensor multiplet,
and the matter content between $E_6$ and $\SU(3)$ is in fact the rank-1 E-string theory, via the embedding \begin{equation}
E_6 \times \SU(3) \subset E_8.
\end{equation}

The anomalies of two rank1 E-string theories and a SU(3) vector multiplet plus one (1,0) tensor multiplet is given by
\begin{align}
&I^{\text{one-loop}}   \nonumber \\
=&I^\text{rank 1}_\text{E-string} (\Tr F_L^2 + \Tr F_{\SU(3)}^2) + 
I^\text{vec}_{\SU(3)}(\Tr F_{\SU(3)}^2) +  I^\text{tensor}+
I^\text{rank 1}_\text{E-string} ( \Tr F_{\SU(3)}^2 + \Tr F_{R}^2) \nonumber \\
=&\frac1{32}(\Tr F^2_{L})^2  + \frac1{32} (\Tr F^2_{R})^2  + \left( \Tr F^2_{L} + \Tr F^2_{R} \right) \left( \frac1{16} p_1(T) - \frac14 c_2(R) \right)  \nonumber \\
&+ \frac{19}{24}c_2^2(R) -\frac{29}{48}c_2(R)p_1(T) + \frac{373}{5760}p_1^2(T) -\frac{79}{1440}p_2(T) \nonumber \\
&-\frac1{32}(\Tr F^2_{\SU(3)})^2 + \Tr F^2_{\SU(3)} \left( -\frac54 c_2(R) + \frac1{16}p_1(T) + \frac1{16} \Tr F^2_L + \frac1{16} \Tr F^2_R \right) \nonumber
\end{align}
where $F_L$, $F_{\SU(3)}$ and $F_R$ are background field strength of $E_6^L$, SU(3) and $E_6^R$, respectively.  
The anomaly of the rank-1 E-string $I^\text{rank 1}_\text{E-string}$ is given in \eqref{E-string-anomaly}, but note that one needs to subtract the contribution from a free hypermultiplet given in \eqref{freehyper}. 
Also note that we call the these contributions the `one-loop' contribution from the lack of better terminology, although there is no concept of loop computations in the E-string theory.

The Green-Schwarz term which cancels the SU(3) part of the anomalies is found to be 
\begin{equation}
I^{\text{GS}} = \frac12 \left(\frac14 \Tr F^2_{\SU(3)} + 5 c_2(R) - \frac14p_1(T) - \frac14 \Tr F^2_L - \frac14 \Tr F^2_R \right)^2.
\end{equation}

Therefore, the total anomalies is
\begin{multline}
	I^{\text{bif}}_{E_6,E_6}(F_L, F_R) = I^{\text{one-loop}} + I^{\text{GS}} = 
\frac1{16}(\Tr F^2_{L})^2 + \frac1{16}\Tr F^2_{L} \Tr F^2_{R} \\
+ \frac1{16} (\Tr F^2_{R})^2 
+ \left( \Tr F^2_{L} + \Tr F^2_{R} \right) \left( \frac18 p_1(T) - \frac32 c_2(R) \right)  \\
+ \frac{319}{24}c_2^2(R) -\frac{89}{48}c_2(R)p_1(T) + \frac{553}{5760}p_1^2(T) -\frac{79}{1440}p_2(T).
\end{multline}

\paragraph{($E_7$, $E_7$) conformal matter.}
The tensor branch is five dimensional, and the M5-brane fractionates into six.  The structure is given by  \begin{equation}
E_7 |  \text{empty} | \SU(2) | \SO(7) | \SU(2) | \text{empty} | E_7.
\end{equation} To compute the anomaly, we make two pairs coalesce to the situation \begin{equation}
E_7 || \SU(2) | \SO(7) | \SU(2) || E_7.
\end{equation} Now we have  vector multiplets in $\SU(2)\times \SO(7)\times \SU(2)$ plus three $(1,0)$ tensor multiplets.
The matter content between $E_7$ and $\SU(2)$ is again the rank-1 E-string theory, via the embedding \begin{equation}
E_7 \times \SU(2) \subset E_8,
\end{equation} and that between $\SU(2)$ and $\SO(7)$ is the half-hypermultiplet in the fundamental of $\SU(2)$ and the spinor of $\SO(7)$.  

\paragraph{($E_8$, $E_8$) conformal matter.}
The tensor branch is eleven dimensional, and the M5-brane fractionates into twelve.  The structure is given by  \begin{equation}
E_8 |  \text{empty} | \text{empty} |  \SU(2) | G_2 |\text{empty} | F_4 | \text{empty} | G_2 | \SU(2) | \text{empty} | \text{empty} | E_8.\label{foo}
\end{equation} 
The matter content between $\SU(2)$ and $G_2$ is the half-hypermultiplet in the bifundamental.
Each $\SU(2)$ also has a half-hypermultiplet in the fundamental. After coalescing, 
the matter between $G_2\times F_4$ is  again the rank-1 E-string theory, via the embedding \begin{equation}
G_2 \times F_4 \subset E_8.
\end{equation} 

To compute the anomaly, we go to the point where we have  \begin{equation}
E_8 |||  \SU(2) | G_2 || F_4 || G_2 | \SU(2) ||| E_8.
\end{equation}   From the F-theory description given in \cite{Heckman:2013pva}, we see that the matter between $E_8$ and  $\SU(2)$ is now the rank-2 E-string theory, whose anomaly was given as  \eqref{E-string-anomaly} minus \eqref{freehyper}.
Here the $\SU(2)$ gauge group is coupled to the $\SU(2)_L$ symmetry explained  in Sec.~\ref{sec:Est}.
This interpretation can be supported as follows: on a generic point on the tensor branch of the rank-2 E-string, there is one free hypermultiplet, which describes the relative position of 2 M5-branes parallel to the end-of-the-world $E_8$ brane.  This counts as one half-hypermultiplet in the fundamental of $\SU(2)$, which should be identified as the half-hypermultiplet of $\SU(2)$ mentioned just below \eqref{foo}.

\paragraph{General results.}
By doing the same exercise we did in the $E_6$ case for all $(G, G)$ conformal matters, where $G$ is an ADE group, we get the following anomaly polynomial
\begin{multline}
I_{G,G}^\text{bif}(F_L,F_R) = \frac{\alpha}{24} c_2(R)^2 - \frac{\beta}{48} c_2(R)p_1(T) + \gamma \frac{7p_1(T)^2-4p_2(T)}{5760}   \\
+\left( -\frac{x}{8} c_2(R) + \frac{y}{96} p_1(T) \right) (\Tr F^2_L + \Tr F^2_R) \\
+ \frac1{48}\left( \tr _{G} F_L^4 + \tr_{G}F_R^4 \right) -\frac12 \left( \frac14 \Tr F^2_L - \frac14 \Tr F^2_R \right)^2
\end{multline}
where coefficients are listed in Table \ref{tab:mattercoef}.
 From this table, we can easily read off that $\gamma = \text{dim}_{G} + 1$, $x=|\Gamma_G| -  h^\vee_G$ and $y=h^\vee_G$. $\alpha$ and $\beta$ are more complicated combinations of group theoretical data, which we will display as a part of the formula for a general number $Q$ of M5-branes on the ALE singularity below. 

\begin{table}[t]
	\centering
	\begin{tabular}{|c||c|c|c|c|c|}
		\hline
		$G$ &$ \SU(k) $&$ \SO(2k) $&$E_6$&$E_7$&$E_8$\\
		\hline\hline
		$\alpha$& 0 & $10k^2 -57k +81$ & 319 & 1670 & 12489  \\
		\hline
		$\beta$ &0 & $2k^2-3k-9$ & 89 & 250 & 831 \\
		\hline
		$\gamma$&$k^2$ & $k(2k-1) + 1 $  & 79 & 134 & 249 \\
		\hline
		$x$ & 0 & $2k-6$ & 12 & 30 & 90 \\
		\hline
		$y$ & $k$ & $2k-2$ & 12 & 18 & 30 \\
		\hline
	\end{tabular}
	\caption{Table of anomaly coefficients for $(G,G)$ conformal matters.}
		\label{tab:mattercoef}
\end{table}

\subsubsection{Anomaly polynomial}\label{sec:bosh}
Now let us determine the anomaly polynomial of $Q$ full M5-branes on the ALE singularity $\bC^2/\Gamma$. We go to a point on the tensor branch, where it is a quiver gauge theory with flavor and gauge groups $[G_0]\times G_1\times \cdots \times G_{Q-1} \times [G_Q]$. 
We have just computed the anomaly of the ``conformal matters'' of $G_i\times G_{i+1}$. 
We also have $Q-1$ free tensor multiplets, describing the relative positions of the M5-branes. 
In this section we are going to compute the total anomaly. 
We include the center-of-mass motion of $Q$ M5-branes just for convenience of computation, but this does not affect the final result
as long as we subtract the contribution of the center-of-mass mode (both one-loop and Green-Schwarz) at the end of the computation.\footnote{
If we compute the anomaly by the inflow argument as in Appendix~\ref{sec:inflow}, the center of mass mode is automatically included there. }

The one-loop anomaly is then given by
\begin{equation}
I^{\text{one-loop}}=\sum_{i=0}^{Q-1} I^\text{bif}_{G,G}(F_i,F_{i+1})+\sum_{i=1}^{Q-1} I^\text{vec}_G(F_i) +QI^\text{tensor} .
\end{equation}
We find that the gauge anomalies can be canceled by the Green-Schwarz term \begin{equation}
I^{\text{GS}}=\frac12 \sum_{i=0}^{Q-1} {\cal I}_i {\cal I}_i
\end{equation} for the self-dual tensor fields with the Bianchi identity \begin{equation}
d\mathcal{H}_i = {\cal I}_i =  \frac{1}{4} \Tr F_i^2- \frac{1}{4} \Tr F_{i+1}^2 + \frac{1}{2}(2i- Q+1) |\Gamma | c_2(R),   \label{eq:calH}
\end{equation}
where $\mathcal{H}_i~(i=0,1,\cdots,Q-1)$ are the three-form fields of the tensor multiplets whose scalars represent the positions of $Q$ M5-branes.
Combining all of them, we get the anomaly polynomial of $Q$ M5 branes at the ALE singularity $\bC^2/\Gamma_G$: 
\begin{multline}
I^{\text{tot}}_{G} = I^{\text{GS}}+I^{\text{one-loop}} 
= |\Gamma|^2 Q^3 \frac{c_2^2(R)}{24}  
-\frac{Q}{48} c_2(R) \biggl{(}|\Gamma| (r_G +1) -1\biggr{)} \biggl{(}4c_2(R)+p_1(T)\biggr{)} \\
-\frac{Q}8 |\Gamma| c_2(R)(\Tr F_0^2+\Tr F_Q^2) + 
\frac{Q}{8}\biggl{(} \frac16 c_2(R) p_1(T) -\frac16 p_2(T) + \frac1{24} p^2_1(T) \biggr{)} \\
-\frac12 I^\text{vec}(F_0) -\frac12 I^\text{vec}(F_Q). \label{eq:anomalyADE}
\end{multline}

Here we give two comments about the result \eqref{eq:anomalyADE}. The first comment is about the center of mass tensor multiplet. The anomaly polynomial of the UV SCFT is determined by subtracting the contributions of the center of mass tensor multiplet,
\begin{equation}
I^{\text{tot}}=I^{\text{SCFT}}+I^{\text{ten}}+\frac1{2Q}\left( \frac14 \Tr F^2_0 -\frac14 \Tr F_Q^2\right)^2.
\end{equation}
Here the third term is a Green-Schwarz term for the center of mass tensor multiplet: it has the Bianchi identity \begin{equation}
d(\frac{1}{Q}\sum_i {\cal H}_i) = \frac{1}{Q}\sum_i {\cal I}_i =\frac{1}{Q} \left( \frac{1}{4} \Tr F^2_0 -\frac{1}{4} \Tr F_Q^2 \right),
\end{equation} and the additional factor $Q$ comes from the factor $Q$ in front of the kinetic term of the center-of-mass tensor multiplet,
i.e. $\Omega^{\rm center-of-mass}=Q$.

The second comment is about the leading behavior. This field theoretical method gives the cubic behavior $\frac1{24}Q^3 |\Gamma|^2c_2(R)^2$. The coefficient is exactly what is expected from $\text{AdS}_7/\text{CFT}_6$. In fact, the whole structure of \eqref{eq:anomalyADE}, including its coefficients, can be reproduced from an anomaly inflow, as will be explained in Appendix \ref{sec:inflow}.

\subsection{Green-Schwarz terms for F-theory constructions}\label{sec:Fth}
In this subsection we investigate the Green-Schwarz terms for general 6d \Nequals{(1,0)} theories
constructed in \cite{Heckman:2013pva}.
Although the Green-Schwarz contribution for such theories can be computed by the method we have developed so far,
here we want to investigate more direct way to relate the Green-Schwarz terms and F-theory constructions.

\subsubsection{On generic points on the tensor branch}

First, we recall how we can determine the Green-Schwarz terms associated to metric, gauge, and flavor background
starting from Type IIB supergravity on $\mathbb{R}^{1,5}\times B$ where $B$ is a noncompact (possibly singular) manifold 
which contains compact or non-compact rational curves $C_a$ possibly wrapped by 7-branes.
We let  the index $a,b,\cdots$ run through all curves in $B$, while $i,j,\cdots$ are only for compact ones.

The 5-form field strength $F_5$ in 10d spacetime decomposes into 6d  self-dual 3-form field strengths $H_i$ as
\begin{align}
	F_5 = H_i \wedge \omega^i
\end{align}
where $\omega_i$ is the Poincar\'e dual of $C_i$.
If we have the Bianchi identity for the 5-form field strength $F_5$ written as
\begin{align}
	\mathrm{d}F_5=Z,
\end{align}
for some 6-form $Z$ consisting of background field strengths. 
Then the Bianchi identities for the 3-form strengths $H^i$ become
\begin{align}
	\mathrm{d}H_i= I_i,\quad  \eta^{ij}I_j=-\int_{B} Z\wedge  \omega^i,
\end{align}
where the matrix $\eta^{ij} = - \int_{B} \omega^i\wedge\omega^j=-C_i\cdot C_j$ is $-1$ times the intersection form of compact cycles in $B$. 
We extend this intersection form to $\eta^{ia}$, which includes intersections between compact and non-compact cycles.
The contributions for the anomaly polynomial from these Green-Schwarz terms are\footnote{
We use a convention that $F_5$ is anti-self-dual so that the 6d fields $H_i$ become self-dual, because $\omega^i$, which have a negative definite intersection matrix, 
are anti-self-dual.
The minus sign in front of $\frac{1}{2} \int Z^2$ comes from this anti-self-dual (instead of self-dual) property of $F_5$.
}
\begin{align}
	I^\text{GS}=-\frac{1}{2}\int_B Z^2=\frac{1}{2}\eta^{ij}I_i I_j.
	\label{eq:GSYY}
\end{align}
The matrix $\Omega^{ij}$ introduced in \eqref{6dpairing} is given by $\Omega^{ij}=\eta^{ij}$ in this class of theories.

As described in \cite{Sadov:1996zm}, the 10d Green-Schwarz term $Z$ is 
\begin{align}
	Z =\frac{1}{4} c_1(B)\wedge p_1(T) + \frac{1}{4}\sum_a \omega^a \Tr F^2_a 
\end{align}
where $F_a$ is the field strength on the 7-branes wrapping $C_a$.
So we get
\begin{align}
	\eta^{ij}I_j = \frac{1}{4} (\eta^{ia}\Tr F_a^2-K^ip_1(T)),\quad
	K^i:=\int_B c_1(B)\wedge \omega^i = 2-\eta^{ii},
	\label{eq:Y1}
\end{align}
up to the term proportional to $c_2(R)$.
This expression is supposed to be true even for non-perturbative F-theory background,
and can be checked for concrete 6d \Nequals{(1,0)} theories by the method developed throughout this paper.

Next, we consider the terms associated to the $\SU(2)_R$ $R$-symmetry, which are not  directly visible by the geometry of F-theory construction. Each Green-Schwarz term $I_i$ should contain contributions proportional to $c_2(R)$ to cancel mixed gauge-$\SU(2)_R$ anomalies, so we write 
\begin{align}
	\eta^{ij} I_j = \frac{1}{4} (\eta^{ia}\Tr F_a^2-K^ip_1(T))+y^i c_2(R).\label{eq:FtheoryCS}
\end{align}
Our next task is to determine the coefficients $y^i$.
The contribution to the mixed anomalies between gauge and R symmetries from the Green-Schwarz terms are
\begin{align}
	I^\text{GS}  \supset \frac{1}{4} y^i\Tr F_i^2 c_2(R).
\end{align}

Consider a generic point of the tensor branch. There we have only Lagrangian degrees of freedom, that are vector multiplets, hyper multiplets, and tensor multiplets, and only the vector multiplets have mixed anomaly between R and gauge symmetries described in \eqref{intro:gauge}.
Then, if a cycle $C_i$ has a nontrivial gauge group $G_i$,
we can immediately conclude that $y^i=h^\vee_{G_i}$ for that cycle.
This agrees with what we saw in \eqref{eq:rank1}.

For $-1$ and $-2$ curves without gauge groups, we cannot determine $y^i$ with this method.
One can circumvent this problem by going to the points of tensor branch where such curves are shrunk giving rank 1 or 2 E-string theories, 
as we have done in previous subsections. 
We will discuss this process of shrinking curves in the next subsection.
It will turn out that $y^i=1$ gives the consistent results in the process.

Alternatively, $y^i$'s should be fixed so that $I^\text{GS}$ reproduces the correct $Q^3$ dependence of the anomaly polynomials of rank $Q$ \Nequals{(2,0)} or E-string theories. This requires $y^i$ to be $1$ for those curves of self-intersection $-1$ and $-2$, assuming that $y^i$ is independent of 
the information of any other curves $C_j$ for $j \neq i$. Then the subleading terms of $Q$ are also correctly reproduced.

Therefore, we claim that $y^i=1$ for the cycles without gauge groups when none of the cycles are shrunk. 
Then we can calculate the anomaly polynomial for any of 6d \Nequals{(1,0)} theories constructed in \cite{Heckman:2013pva}.
At a generic point of the tensor branch, we have only fields described by Lagrangians, so the calculation of one-loop anomaly polynomial is straightforward. 
Then all we have to do is just add the Green-Schwarz contribution \eqref{eq:GSYY}.

\subsubsection{Shrinking $-1$ curves}
Now let us describe a convenient algorithm for calculating Green-Schwarz terms for a non-generic point of the tensor branch where some of $-1$ curves are shrunk.
This will justify
the above claim that $y^i=1$ for the cycles without gauge groups, and  
also can be used to obtain \eqref{eq:calH}. 
For simplicity, We consider blowing-down a certain $-1$ curve $C_A$ in $B$
which intersect with curves $C_{A-1}$ and $C_{A+1}$.
We can obtain the result for the case multiple $-1$ are shrunk by recursion.

Let $\widehat{B}$ be the manifold obtained by shrinking (i.e., blowing down) the $-1$ curve $C_A$ in $B$, and $p:B\to\widehat{B}$ be the blow-down map.
The homology classes of cycles $\widehat{C}_i=p(C_i),i\neq A$ in $\widehat{B}$ and $C_i$ in $B$ are related by
\begin{align}
	p^*[\widehat{C}_i]=
	\begin{cases}
		[C_i]+[C_A] & i=A-1,A+1\\
		[C_i] & i\neq A-1,A,A-1,
	\end{cases}
\end{align}
where $[C_i]$ means the homology class of $C_i$.
In the following, we demand that indices $i,j$ do not take the value $A$.
Their intersection form becomes
\begin{align}
	\widehat{\eta}^{ij} &=-\widehat{C}^i\cdot\widehat{C}^j 
	=
	\begin{cases}
		\eta^{ii}-1 & i=j=A\pm1 \\
		-1	& (i,j)=(A-1,A+1),(A+1,A-1)\\
		\eta^{ij} & \text{otherwise}
	\end{cases}.
\end{align}
There are 3-form field strengths $\widehat{H}_i$ each associated to cycles $\widehat{C}_i$ which satisfy
\begin{align}
	\mathrm{d}\widehat{H}_i&=\widehat{I}_i,\nonumber\\
	\widehat{I}^i &=
	\begin{cases}
		 I^i+I^A & i=A+1,A-1\\
		 I^i & \text{otherwise}
	\end{cases}.\label{eq:hatYY}
\end{align}
where $I^i=\eta^{ij} I_j$.
This means that 
the $\Tr F_i^2$ and $p_1(T)$ dependence of the Green-Schwarz terms $\widehat{I}^i$
are again written by \eqref{eq:Y1}, as it should be.  
 The coefficients of $c_2(R)$ in $\widehat{I}^i$
can be easily calculated from \eqref{eq:hatYY}.

Then, the new Green-Schwarz contribution to the anomalies after the blow-down is 
\begin{align}
	\widehat{I}^\text{GS} &= \frac{1}{2}\widehat{\eta}^{ij}\hat{I}_i\hat{I}_j=\frac{1}{2}(\widehat{\eta}^{-1})_{ij}\hat{I}^i\hat{I}^j.
	\label{eq:hatY}
\end{align} The anomaly of the total system is \begin{equation}
I^\text{tot}= I^\text{one-loop} + I^\text{GS} = \widehat I^\text{one-loop}+ \widehat I^\text{GS}.
\end{equation} This equality must hold because of the anomaly matching. We also have a mathematical relation\footnote{
The relation is shown as follows. Let us change the basis of two-forms of $B$ from $\omega^i$ to 
${\omega}'^i=p^*\hat{\omega}^i$ $(i \neq A)$, $\omega'^A=\omega^A$.
Then, $I'^i=-\int Z \wedge \omega'^i$ is given by $I'^i=\widehat{I}^i$  ($i \neq A$) and $I'^A=I^A$. The ``intersection form" in the new basis 
$\eta'^{ij}=-\int \omega'^i \wedge \omega'^j$ is given by a block diagonal form, $\eta'^{ij}=\widehat{\eta}^{ij}$ $(i,j \neq A)$, 
$\eta'^{iA}=0$ $(i \neq A)$ and $\eta'^{AA}=1$. Using these, \eqref{eq:amatheq} immediately follows.}
\begin{align}
	I^\text{GS}=\widehat{I}^\text{GS}+\frac{1}{2}(I^A)^2. \label{eq:amatheq}
\end{align} Thus the one-loop factor before and after the blow-up must be related by 
\begin{align}
	\widehat{I}^\text{one-loop}=I^\text{one-loop}+\frac{1}{2}(I^A)^2.\label{eq:UVIRoneloop}
\end{align}

Let us apply the result \eqref{eq:UVIRoneloop} to the case where the curve $C_A$ does not have any gauge group.
Then a copy of the rank-1 E-string theory appears by the blow-down.
So the difference between the one-loop anomalies before and after the blow-down
must be the difference between the rank-1 E-string and a free tensor multiplet,
\beq
\widehat{I}^\text{one-loop}-I^\text{one-loop}=I^{\rm rank1}_{\rm E-string}-I^{\rm tensor}=\frac{1}{2} \left(c_2(R)-\frac{1}{4}p_1(T)-\frac{1}{4} \Tr F^2_{E_8} \right)^2.
\eeq
On the other hand, from \eqref{eq:FtheoryCS} we get
\beq
\frac{1}{2}(I^A)^2=\frac{1}{2} \left(y^A c_2(R)-\frac{1}{4}p_1(T) -\frac{1}{4}\Tr F^2_{A-1}-\frac{1}{4}\Tr F^2_{A+1} \right)^2.
\eeq
For these results to be consistent, we must have $y^A=1$, justifying our claim in the previous subsection.
We can also see that the gauge groups $G_{A-1} \times G_{A+1}$ are embedded in the $E_8$ of the E-string theory
such that $\Tr F^2_{E_8}=\Tr F^2_{A-1}+\Tr F^2_{A+1}$.

As an example, let us calculate the Green-Schwarz terms associated to the tensor branch mode which represents the distance between 
two full M5-branes on $E_6$ type singularity using \eqref{eq:hatYY}. In the dual F-theory description,
the intersections of cycles and the coefficients $y_i$'s at a generic point of the tensor branch (corresponding to fractionated M5-branes) are given by

\begin{center}
\begin{tabular}{cccccccccc}
& \fbox{$E_6$}  & $|$  empty  $|$&SU(3) &$|$ empty $|$&  $E_6$ &$|$ empty $|$ & SU(3) & $|$ empty $|$&  \fbox{$E_6$}\\
 $\eta^{ii}$ &        & 1         &      3     &    1 &  6   &    1     &  3      &  1 \\
 $y^i$         &       & 1          &        3    & 1    & 12 &    1     & 3      &   1\\
 \end{tabular}
\end{center}

Here the spaces between two adjacent $|$ and $|$ represent compact cycles and we explicitly write the corresponding gauge symmetries as $E_6$, $\SU(3)$ or empty. The leftmost and rightmost \fbox{$E_6$} represent the noncompact cycles which provide $E_6 \times E_6$ flavor symmetries. The numbers  in the second and third rows are $\eta^{ii}$ (i.e. $-1$ times the self-intersection number of the corresponding compact cycle) and the coefficients $y^i$, respectively.

When we shrink all of the $-1$ curves in the above figure, we get 
\begin{center}
\begin{tabular}{cccccc}
& \fbox{$E_6$}  & $||$ SU(3) $||$&  $E_6$ & $||$ SU(3) $||$ &  \fbox{$E_6$}\\
 $\eta^{ii}$ &        & 1         &      4     &    1  \\
 $y^i$         &       & 5          &        14    & 5\\
 \end{tabular}
\end{center}
 where $||$ represents that the curve in-between has shrunken, and finally this goes to
\begin{center}
\begin{tabular}{cccc}
& \fbox{$E_6$}  & $||||$ $E_6$ $||||$&  \fbox{$E_6$}\\
 $\eta^{ii}$ &        & 2   \\
 $y^i$         &       & 24 \\
 \end{tabular}
\end{center}
The symbol $||||$ now corresponds to a full M5 brane in M-theory, and therefore the tensor mode between two $||||$ is what we wanted.
Note that $24$ is equal to their order $|\Gamma_{E_6}|$ of the binary tetrahedral group $\Gamma_{E_6}$. 

In general, the F-theory description of $Q$ separated (but not fractionated) full M5-branes on the $\bC^2/\Gamma$ singularity 
is a sequence of $Q-1$ curves of self-intersection number $-2$ which are decorated by the gauge group $G$.
Let $H_i$ be the tensor multiplet for the $i$-th 2-cycle, with the Bianchi identity $dH_i=I_i$.
This $I_i$ can be easily determined  in the same way as the above computation for $E_6$, and we obtain
\begin{align}
	I^i=\eta^{ij}I_j=\frac{1}{4}(2\Tr F_i^2-\Tr F_{i-1}^2 -\Tr F_{i+1}^2 )+|\Gamma|c_2(R).
\end{align}
Note that the tensor multiplet containing $H^i=\eta^{ij}H_j$ corresponds to the distance of two M5-branes.\footnote{
The scalars $\phi_i$ in the tensor multiplets may be contained in the K\"ahler form $J$ of the base $B$ as $J=-\phi_i \omega^i+\cdots$.
The areas of the cycles $C_i$ are given by $\int_{C_i}J=\int_B \omega^i \wedge J=\eta^{ij} \phi_j=\phi^i$, and they corresponds to the distances 
between adjacent (full or fractional) M5-branes.}
On the other hand, the positions of the M5-branes
are denoted by $\mathcal{H}_{i}$ in Sec.~\ref{sec:bosh}, and hence we have~\begin{equation}
H^i=\mathcal{H}_{i}-\mathcal{H}_{i-1}.
\end{equation} This means that $I^i$ above should be given by ${\mathcal I}_i-{\mathcal I}_{i-1}$, where $d\mathcal{H}_i={\cal I}_i$ was given in \eqref{eq:calH}, and this is indeed the case.

\section{Conclusions and discussions}\label{sec:conclusions}
In this paper, we described methods that allow us to determine the anomaly polynomials of very general 6d \Nequals{(1,0)} superconformal theories.  
The essential idea was that the tensor branch vevs do not break any symmetry other than the conformal symmetry, and therefore the anomaly polynomial at the origin of the tensor branch can be obtained by the anomaly matching on the tensor branch. 
For this, we need to determine the Green-Schwarz term carried by the self-dual tensor fields on the tensor branch.   
We described two methods to do so. 

The first was applicable when there was no gauge field on generic points on the tensor branch. In this case, we had sufficient control of the behavior of the $S^1$ compactification. Then we can determine the induced Chern-Simons terms in 5d, that can then be lifted to 6d to fix the Green-Schwarz term.
The second was applicable when the number of the gauge fields and the number of the tensor fields are equal on some points on the tensor branch.  
In this case,  we can determine the Green-Schwarz term just by requiring that there is no gauge anomaly. 

We used the first method to derive the anomaly polynomials of \Nequals{(2,0)}  theories of arbitrary type, and of the E-string theory of arbitrary rank.  In most of the cases, the results were known from the analysis of the M-theory anomaly inflow, and our method gives an independent confirmation. For \Nequals{(2,0)} theory of type $E$, ours is the first derivation. 

We then used the second method to derive the anomaly polynomials of the worldvolume theories on $Q$ M5-branes on the ALE singularities. We found a general formula, that can be successfully checked against the anomaly inflow computation reported in Appendix~\ref{sec:inflow}. We also gave a general procedure to determine the Green-Schwarz contribution for \Nequals{(1,0)} theories recently discussed using F-theory in \cite{Heckman:2013pva,DelZotto:2014hpa}. 

Our methods can definitely be used to study various other \Nequals{(1,0)} theories already known in the literature. Hopefully, the anomaly polynomials we determined here and the methods themselves would be useful in the study of the compactifications of \Nequals{(1,0)} theories to various lower dimensions. 

In this paper, we only considered the part of the anomalies of the \Nequals{(1,0)} theories that can be captured at the level of differential forms. It would be interesting to study the global anomalies of these theories, following \cite{Monnier:2013rpa,Monnier:2014txa}. Also, the partition functions of \Nequals{(2,0)} theories are known to behave rather like conformal blocks of two-dimensional chiral CFTs  \cite{Witten:1998wy,Witten:2009at,Henningson:2010rc,Tachikawa:2013hya}, and it would be interesting to understand what happens in \Nequals{(1,0)} cases. 

Another possible application of our results is the following.
In the case of 4d SCFTs, there are relations between the coefficients of anomaly polynomials and central charges
$a,c$ and other flavor central charges~\cite{Anselmi:1997am,Anselmi:1997ys}.
If there are similar relations also in 6d SCFTs, the anomaly polynomials may be used to calculate the central charges.
In particular, it might be useful for checking whether the $a$-theorem in 6d is valid or not.
(See \cite{Cardy:1988cwa,Myers:2010tj,Elvang:2012st,Yonekura:2012kb,Grinstein:2014xba} for some evidences for or against the $a$-theorem in 6d).
One observation is that the Green-Schwarz contribution $\Omega^{ij}I_i I_j$ has a certain positivity property because 
the matrix $\Omega^{ij}$ is positive-definite. If a UV SCFT flows to an IR SCFT and some free fields including tensor multiplets,
the coefficient of e.g. $c_2(R)^2$ always decreases between the UV and IR SCFT, assuming that the UV and IR $\SU(2)_R$ are the same. 
Therefore, by relating the coefficients of the anomaly polynomials
to $a$, we may be able to have an evidence for the 6d $a$-theorem.

\section*{Acknowledgments}
YT thanks K. Intriligator for discussions.
KO and HS are partially supported by the Programs for Leading Graduate Schools, MEXT, Japan,
via the Advanced Leading Graduate Course for Photon Science
and via the  Leading Graduate Course for Frontiers of Mathematical Sciences and Physics, respectively. 
KO is also supported by JSPS Research Fellowship for Young Scientists.
YT is  supported in part by JSPS Grant-in-Aid for Scientific Research No. 25870159,
and in part by WPI Initiative, MEXT, Japan at IPMU, the University of Tokyo.
The work of KY is supported in part by NSF Grant PHY-0969448.
 
\appendix

\section{Tables of anomalies and group theoretic constants}\label{sec:tables}
In this Appendix we summarize the anomaly polynomials for multiplets of 6d \Nequals{(1,0)} supersymmetry, and other group theoretic notations. In this paper we do not concern about subtleties arise from global structures of gauge groups and be careless about whether we are talking about groups or algebras.

In this paper we use the notation in which the anomaly polynomial of Weyl fermions in a representation $\rho$ becomes
\begin{align}
	\hat{A}(T)\mathrm{tr}_\rho\mathrm{e}^{\mathrm{i}F}.
\end{align}
where $\hat{A}(T)$ is the A-roof genus.
In particular, $F$ is anti-Hermitican and include a $(2\pi)^{-1}$ factor in its definition compared to the usual one.
The anomaly polynimials for \Nequals{(1,0)} multiplets are the following:
\begin{itemize}
	\item Hypermultiplet with representation $\rho$
		\begin{align}
			\frac{\mathrm{tr}_\rho F^4}{24}+\frac{\mathrm{tr}_\rho F^2 p_1(T)}{48}+d_\rho\frac{7p_1^2(T)-4p_2(T)}{5760}
		\end{align}
	\item Vector multiplet with group $G$
		\begin{align}
			-\frac{\mathrm{tr}_{\mathrm{adj}}F^4+6 c_2(R) \mathrm{tr}_{\mathrm{adj}}F^2+d_{G}c_2(R)^2}{24}&-\frac{(\mathrm{tr}_{\mathrm{adj}} F^2 + d_{G}c_2(R)) p_1(T)}{48}
			\nonumber\\
			&-d_G\frac{7p_1^2(T)-4p_2(T)}{5760}
		\end{align}
	\item Tensor multiplet
		\begin{align}
			\frac{c_2(R)^2}{24}+\frac{c_2(R)p_1(T)}{48}+\frac{23 p_1(T)^2-116 p_2(T)}{5760}
		\end{align}
\end{itemize}
where $d_{\rho}$ and $d_{G}$ are the dimensions of representation $\rho$ and group $G$, respectively.

It is convenient to define the symbol $\Tr _{G}$ to be the trace in the adjoint representation divided by the dual Coxeter number $h^\vee_{G}$ of the gauge group $G$, listed in Table \ref{tab:groupconst}.
\begin{table}[t]
	\centering
	\begin{tabular}{|c||c|c|c|c|c|c|c|c|}
		\hline
		$G$ &$ \SU(k) $&$ \SO(k) $&$\USp(2k)$&$G_2 $&$F_4 $&$E_6$&$E_7$&$E_8$\\
		\hline\hline
		$r_{G}$&$k-1$ & $\lfloor k/2 \rfloor$ & $k$ & 2 & 4 & 6 & 7 & 8\\
		\hline
		$h^\vee_{G}$&$k$ & $k-2$ & $k+1$ & 4 & 9 & 12 & 18 & 30\\
		\hline
		$d_{G}$&$k^2-1$ & $k(k-1)/2$ & $k(2k+1)$ & 14 & 52 & 78 & 133 & 248\\
		\hline
		$d_{\mathrm{fnd}}$&$k$ & $k$ & $2k$ & 7 & 26 & 27 & 56 & 248\\
		\hline
		$s_{G}$&$\frac12$ & $1$ & $\frac12$ & 1 & 3 & 3 & 6 & 30\\
		\hline
		$t_{G}$&$2k$ & $k-8$ & $2k+8$ & 0 & 0 & 0 & 0 & 0\\
		\hline
		$u_{G}$&$2$ & $4$ & $1$ & $\frac{10}{3} $&$ 5 $&$ 6 $&$ 8$&$ 12$\\
		\hline
	\end{tabular}
	\caption{Group theoretical constants defined for all $G$. 
Those constants are also listed in Appendix of \cite{Erler:1993zy}.
		\label{tab:groupconst}}
\end{table}
One of the properties of $\Tr $ is that $\frac{1}{4}\int \Tr F^2$ is one when there is one instanton on a four-manifold.
Moreover, if we have subgroup $G'$  in a  group $ G$ with Dynkin index of embedding $1$, for an element $f$ of universal enveloping algebra of Lie algebra of $G'$ , the following equation holds:
\begin{align}
	\Tr _{G'}f=\Tr _{G}f.
\end{align}
All of the embeddings we consider in this paper have index $1$, so we often omit the subscription $G$ in $\Tr _{G}$.

To convert the above anomaly polynomials to a convenient form, we define some constants and write those values in Table \ref{tab:groupconst}. We define the constant $s_{G}$ which relates the trace of $F^2$ in the fundamental representation and $\Tr F^2$ as $\tr_{\rm fund }F^2=s_{G} \Tr F^2$.
Then we have 
\begin{align}
	\mathrm{tr}_{\mathrm{adj}}F^2&= h^\vee_G \Tr  F^2, &
	\mathrm{tr}_{\mathrm{fund}}F^2&= s_G \Tr  F^2,
\end{align}
where the first equation is just the definition of $\Tr$. For trace of $F^4$, we define $t_G$ and $u_G$ by
\begin{align}
	\mathrm{tr}_{\mathrm{adj}}F^4=t_{G}\mathrm{tr}_{\mathrm{fnd}}F^4+\frac{3}{4}u_G( \Tr F^2)^2.
\end{align}

\begin{table}[t]
	\centering
	\begin{tabular}{|c||c|c|c|c|c|c|c|c|}
		\hline
		$G$ &$ \SU (2) $&$ \SU(3) $ & $G_2 $&$F_4 $&$E_6$&$E_7$&$E_8$\\
		\hline\hline
		$w_{G}$&$\frac83$ & 3 & $\frac{10}3$ & 5 & 6 & 8 & 12\\
		\hline
		$x_{G}$&$\frac1{6}$& $\frac1{6}$ &  $\frac13$ & 1 & 1 & 2 & 12\\
		\hline
	\end{tabular}
	\caption{Group theoretical constants defined only  $G$ without independent quartic Casimir. }
		\label{tab:groupconst2}
\end{table}

For gauge groups $G=\SU(2),\SU(3)$ and all exceptional groups, there are no independent quadratic Casimir operator, so we can relate $\mathrm{tr}_{\mathrm{\rho}}F^4$ and $(\Tr F^{2})^2$
by
\begin{align}
	\mathrm{tr}_{\mathrm{adj}}F^4&= \frac34 w_G  (\Tr F^2)^2,  &
	\mathrm{tr}_{\mathrm{fund}}F^4&= \frac34 x_G (\Tr F^2)^2
\end{align}  
These constants are tabulated in Table~\ref{tab:groupconst2}.
Note that because $t_{\SO(8)}=0$, we can also relate $\tr_{\mathrm{adj}}F^4$ to $(\Tr F^2)^2$ for $G=\SO(8)$.

All representations we use in this paper are fundamental or adjoint, except for the spin representation $\mathbf{8}$ of $\SO(7)$. The conversion constant for this representation is
\begin{align}
	\mathrm{tr}_{\mathbf{8}}F^2&=\Tr F^2,\nonumber\\
	\mathrm{tr}_{\mathbf{8}}F^4&=-\frac{1}{2}\mathrm{tr}_{\mathrm{fund}}F^4+\frac{3}{8}(\Tr F^2)^2.
\end{align}

Finally, let us note that the finite subgroup $\Gamma_G$ of $\SU(2)$ of type $G=A_{n},$ $D_n$ and $E_n$ has the following order: \begin{equation}
|\Gamma_{\SU(k)}|=k,\quad
|\Gamma_{\SO(2k)}|=4k-8,\quad
|\Gamma_{E_6}|=24,\quad
|\Gamma_{E_7}|=48,\quad
|\Gamma_{E_8}|=120.
\end{equation}

\section{Anomaly of M5s on ALE singularity via inflow}\label{sec:inflow}

Here we derive the anomaly polynomials of the theory realized by M5 branes put on the ALE singularities by using the anomaly inflow.
What we will compute includes contributions not only from the genuine SCFT part, but also from the center of mass tensor multiplet and its Green-Schwarz contribution. 

\subsection{Chern-Simons terms in M-theory}
For  the anomaly inflow, Chern-Simons terms involving the M-theory three-form $C$ is important. 
In the eleven dimensional spacetime $X_{11}$, if there is no magnetic source for the field strength four-form $G=dC$, we have 
\beq
S_{CGG}=\frac{2\pi}{6}& \int_{X_{11}} C \wedge G \wedge G=\frac{2\pi}{6}\int_{Y_{12}} G \wedge G \wedge G, \\
S_{CI_8} =-2\pi & \int_{X_{11}} C \wedge I_8=-2\pi\int_{Y_{12}} G \wedge I_8, 
\eeq
where in the following, $Y_{p+1}$ means a $p+1$ dimensional manifold whose boundary is $X_{p}$, i.e., $\partial Y_{p+1}=X_{p}$,
and 
\beq
I_8 =\frac{1}{48} \left[  p_2(TX_{11}) -\frac{1}{4}p_1^2(TX_{11}) \right].
\eeq

When there is an orbifold singularity $X_{11} =X_7 \times {\bC^2/ \Gamma}$, we also have two types of Chern-Simons terms localized on the singularity.
The first one can be determined in the following way. When $X_{11} =X_7 \times {\bC^2/ \Gamma}$, the structure group of the tangent bundle
is decomposed as $\SO(11) \to \SO(7) \times \SU(2)_L \times \SU(2)_R$. The orbifold $\Gamma$ acts on $\SU(2)_L$.
There is an $\SU(2)$ symmetry acting on $\bC^2/ \Gamma$, and by a slight abuse of notation, we denote this symmetry as $\SU(2)_R$. When $\Gamma$ is of type $A_k$, there is a $\U(1)$ symmetry acting on $\mathbb{C}^2/\Gamma$, but we ignore this symmetry for simplicity.
Let $c_2(L)$ and $c_2(R)$ be the Chern classes of $ \SU(2)_L $ and $ \SU(2)_R$ respectively. This $c_2(R)$ gives the Chern class
of the connection field associated to the rotational symmetry $\SU(2)_R$.
Then $I_8$ becomes
\beq
I_8=-\frac{1}{48}c_2(L) (4c_2(R)+p_1(TX_7))+\frac{1}{48}\left[p_2(T)-p_1(T)c_2(R)-\frac{1}{4}p_1(T)^2\right].
\eeq
The singularity may be regarded as a gravitational instanton, and has some nontrivial curvature $c_2(L)$ localized at the singularity, with
\beq
\int_{ \bC^2 / \Gamma} c_2(L) =: \chi_{\Gamma}.
\eeq
where $\chi_{ \Gamma}$ can be thought of as  a version of the ``Euler number'' of the singularity. 
Then, we get a Chern-Simons term on the singularity as
\beq
S_{CI_8} =S_{CI_8}^{\rm bulk}+2\pi \int_{X_7 \times \{0\}} \frac{\chi_{\Gamma}}{48} C \wedge (4c_2(R)+p_1(T)) ,
\eeq
where $S_{CI_8}^{\rm bulk}$ is the contribution which is not localized on the singularity.

The value of $\chi_\Gamma$ is given as \cite{Gibbons:1979gd}
\beq
\chi_\Gamma=r_\Gamma+1 -\frac{1}{|\Gamma|},\label{eq:singularEuler}
\eeq
where $r_\Gamma$ is the rank of the $A_r, D_r, E_r$ group corresponding to $\Gamma$, and $|\Gamma|$ is the order of $\Gamma$.

This formula can be understood as follows. Let $M=\{ z \in \bC^2/\Gamma ; |z|^2 \leq 1  \}$.
The boundary is $\partial M=S^3/\Gamma$.
The topological Euler number of this space is $\chi(M)=r+1$, because $\dim H_2(M)=r$, $\dim H_0(M)=1$ and others are zero.
Now recall that the topological Euler number is also given as an integral of local quantities as $\chi(M)=\int_M E_4+\int_{\partial M} ({\rm local~term})$,
where we have denoted the Euler density as $E_4$. When $\Gamma$ is trivial so that $\partial M=S^3$, 
the contribution from the boundary integral is $1$ because
$r=0$ and $E_4=0$ in that case. Then this boundary contribution is $1/|\Gamma|$ when $\partial M=S^3/\Gamma$. 
Therefore we get $\int_{M} E_4=r+1-1/|\Gamma|$.

The second type of Chern-Simons terms involves gauge fields localized on the singularity.
The gauge fields $A_i~(i=1,\cdots,r_\Gamma)$ in the Cartan subalgebra of $A_r, D_r,E_r$ gauge algebra localized on the singularity 
comes from the three-form $C$ as
\beq
C=C^{\rm bulk}+i\sum_{i=1}^r \omega^i \wedge A_i,
\eeq
where $\omega^i$ are Poincare duals to the two cycles which are collapsed at the singularity.
The factor $i=\sqrt{-1}$ was introduced to make $A_i$ anti-Hermitian. 
Then we get
\beq
S_{CGG}&=S_{CGG}^{\rm bulk}+ \frac{2\pi}{2} \eta^{ij} \int_{X_7 \times  \{0\}} C^{\rm bulk} \wedge F_i \wedge F_j \nonumber\\
&= S_{CGG}^{\rm bulk}+ \frac{2\pi}{4}\int_{X_7 \times \{0\}} C^{\rm bulk} \wedge \Tr F^2,
\eeq
where $\eta^{ij}=-\int \omega^i \wedge \omega^j$ is $-1$ times the intersection matrix of the two-cycles given by the Cartan matrix of $A_r, D_r,E_r$.
Although it is obtained for gauge fields in the Cartan subalgebra, the last expression should be valid for more general non-abelian fields.

Combining the above results, we get the Chern-Simons terms localized on the singularity as
\beq
S_{\Gamma}= & 2\pi \int_{X_7 \times \{0\}} C^{\rm bulk} \wedge J_4, \\
J_4 \equiv & \frac{\chi_\Gamma}{48}(4c_2(R)+p_1(T)) +\frac{1}{4} \Tr F^2 .\label{eq:Jpoly}
\eeq

\subsection{Effects of M5 on the theory on singularity}
Before computing the inflow, let us explain the effects of inserting M5 branes on the singularity to the gauge fields living on the singularity. 
Consider the $A$-type singularity, and suppose that this singularity is realized in Taub-Nut space instead of ALE space.
Then, by going to the type IIA description, the singularity becomes D6 branes and the M5 branes become NS5 branes.
We get a system where NS5 branes are inserted to D6. In this case, NS5 branes have the effects that the two sides of the 
NS5 become independent gauge theories, i.e., the gauge group is $\SU(r+1)_L \times \SU(r+1)_R$ where $\SU(r+1)_L$ is on one side of the NS5 brane
and $\SU(r+1)_R$ is on the other. Furthermore, the boundary condition of these gauge fields is such that a gauge theory between two NS5
becomes 6d \Nequals{(1,0)} vector multiplet instead of \Nequals{(1,1)} vector multiplet.
That is, among the 7d \Nequals{1} fields, three scalars and a component of vector field normal to NS5 have Dirichlet boundary conditions,
while vector fields tangent to NS5 have Neumann boundary condition.
The same things should happen when M5 branes are inserted in more general $A,D,E$ singularities.

The above boundary condition gives contribution to the anomaly, just as in the case of the end-of-the-world $E_8$ brane where
the change of the gravitino boundary condition gave contributions to the anomaly~\cite{Horava:1995qa,Horava:1996ma}.
This contribution is given by
\beq
-\frac{1}{2} I^{\rm vec}_L-\frac{1}{2} I^{\rm vec}_R, \label{eq:gauginoeffect}
\eeq
where $I^{\rm vec}_L$ and $I^{\rm vec}_R$ are the anomalies of \Nequals{(1,0)} vector multiplets with gauge groups $G_L$ and $G_R$
on the two sides of the M5 branes, respectively.

\subsection{Anomaly inflow on $\bR \times \bC^2/\Gamma$}
Now we calculate the anomaly inflow. Since the relevant calculations are almost the same as in \cite{Ohmori:2014pca}, 
we will be brief and neglect some of the subtleties. 

Let us take the eleven dimensional space to be $X_{11}=X_6 \times (\bR \times \bC^2/\Gamma)$ and put $Q$ M5 branes at the origin
of $\bR \times \bC^2/\Gamma$. Let $y^a~(a=1,2,3,4,5)$ be the coordinates of the covering space $\bR^5=\bR \times \bC^2$.

If $\Gamma$ is trivial, the Bianchi equation for $G$ is
\beq
dG=Q \prod_{a=1}^5 \delta(y^a) dy^a.
\eeq
Its solution at $y \neq 0$ is given by 
\beq
G=\frac{Q}{2} e_4+({\rm regular}), 
\eeq
where (regular) represents terms that are not singular at $y=0$. 
The four-form $e_4$ which is closed at $y \neq 0$ is given by
\beq
e_4(y)=& \frac{1}{32 \pi^2} \epsilon_{a_1 \cdots a_5}\Big[ (D \hat{y})^{a_1}(D \hat{y})^{a_2}(D \hat{y})^{a_3}(D \hat{y})^{a_4} \hat{y}^{a_5} \nonumber \\
&~~~~~~~~~~~~~~~~~~-2F^{a_1 a_2} (D \hat{y})^{a_3}(D \hat{y})^{a_4}\hat{y}^{a_5}+F^{a_1 a_2}F^{a_3 a_4}\hat{y}^{a_5} \Big],
\eeq
where $\hat{y}^a=y^a/|y|$, $D$ is a covariant exterior derivative of $\SO(5)_R$ rotational symmetry around the origin of $\bR^5$, 
and $F^{a_1 a_2}$ is the field strength of $\SO(5)_R$. 
We restrict the $\SO(5)_R$ bundle to the subbundle $\SU(2)_R \subset \SO(4)_R \subset \SO(5)_R$,
which will also be preserved when the space is divided by $\Gamma$.
Then, when we introduce the orbifold, the only change is that $G=(|\Gamma|Q/2) e_4+({\rm regular})$ as long as 
$y^a$ are understood to be the coordinates of the covering space.

Some of the important properties of $e_4$ we will use are
\beq
\int_{S^4} e_4=2,~~~\int_{S^4} (e_4)^3=2 c_2(R)^2,~~~e_4|_{y^{2,3,4,5}=0}=-c_2(R)\sign(y^1),
\label{eq:e4formula1}
\eeq
where $S^4$ is a sphere around the origin of $\bR \times \bC^2$.
When we divide by $\Gamma$, there is an additional factor $1/|\Gamma|$ in the first two equations.

Now let us determine the contribution from the inflow from the Chern-Simons terms.
Because $G$ is singular at the position of the M5 branes, we remove a small tubular neighborhood 
of the M5 branes in the integral of Chern-Simons terms. We denote a tubular neighborhood of a submanifold $M$ as $D_{\epsilon}(M)$.
By an abuse of notation, we denote the submanifold where the M5 branes are located as $X_6$ (or $Y_7$ depending on whether we consider 
$X_{11}$ or $Y_{12}$).

Because $e_4$ is closed, it is (locally) written as $e_4=d e_3^{(0)}$
Now, the most singular part of the Chern-Simons term $S_{CGG}^{\rm bulk}$ is given as
\beq
S_{CGG}^{\rm bulk}=&\frac{2\pi}{6} \lim_{\epsilon \to 0}\int_{Y_{12} \setminus D_{\epsilon}(Y_7) } 
G^{\rm bulk} \wedge G^{\rm bulk} \wedge G^{\rm bulk} \nonumber \\
\sim & 2\pi \cdot \frac{Q^3 |\Gamma|^3}{48} \lim_{\epsilon \to 0}\int_{Y_{12} \setminus D_{\epsilon}(Y_7) }  e_4^3 
=-2\pi \cdot \frac{Q^3 |\Gamma|^3}{48} \lim_{\epsilon \to 0} \int_{ \partial D_{\epsilon}(Y_7) } e_3^{(0)} e_4^2 \nonumber \\
=&-2\pi \cdot \frac{Q^3 |\Gamma|^2}{24} \int_{ Y_7 } c_2(R)^{(0)} c_2(R),
\eeq
where $dc_2(R)^{(0)} =c_2(R)$ and we have used the second equation of \eqref{eq:e4formula1} in the last equation.
Thus the contribution of this to the anomaly polynomial is $-(Q^3|\Gamma|^2/24)c_2(R)^2$. 

In the same way, we get
\beq
S_{CI_8}^{\rm bulk} \sim & 2\pi \cdot Q  \int_{ Y_7 } I_7^{(0)} ,\nonumber \\
S_{\Gamma} \sim & 2\pi \cdot \frac{Q|\Gamma|}{2} \int_{ Y_7 } c_2(R)^{(0)}  \left( J_{4,L}+J_{4,R} \right),
\eeq
where $dI_7^{(0)}=I_8$, the $J_{4,L}$ and $J_{4,L}$ are the $J_4$ defined in \eqref{eq:Jpoly} on the left and right of the M5 branes respectively,
and we have used the first and third equations of \eqref{eq:e4formula1}.

Combining these, the inflow of anomaly is given by
\beq
-\frac{Q^3|\Gamma|^2}{24}c_2(R)^2+QI_8+\frac{Q|\Gamma|}{2}  c_2(R)  ( J_{4,L}+J_{4,R}).
\eeq
This must be cancelled by the anomaly of the theory living on the M5 branes.
Taking into account the contribution \eqref{eq:gauginoeffect}, we finally get the anomaly polynomial of the theory on M5 branes
which are put on $\bR \times \bC^2/\Gamma$ as
\beq
&I^{\rm tot}(Q~{\rm M5};~\bR \times \bC^2/\Gamma) \nonumber \\
&=\frac{Q^3|\Gamma|^2}{24}c_2(R)^2-QI_8-\frac{Q|\Gamma|}{2} c_2(R)  ( J_{4,L}+J_{4,R})-\frac{1}{2} I^{\rm vec}_L-\frac{1}{2} I^{\rm vec}_R.
\eeq
We can check that this formula is equal equal to \eqref{eq:anomalyADE}.
 
\bibliographystyle{ytphys}
\bibliography{ref}

\end{document}